\documentclass[12pt]{iopart}
\usepackage{graphicx}

\begin{document}

\title[SW defects cause metal-semiconductor transition in CNTs depending on their orientation]{Stone-Wales defects can cause metal-semiconductor transition in carbon nanotubes depending on their orientation}

\author{P. Partovi-Azar$^1$, A. Namiranian$^2$}
\address{$^1$Computational Physical Science Laboratory, Department of Nano-Science, Institute for Research in Fundamental Sciences
(IPM), P.O. Box 19395-5531, Tehran, Iran.}
\address{$^2$Department of Physics, Iran University of Science and Technology, Narmak, 16345 Tehran, Iran.}

\begin{abstract}

It has been shown that the two different orientations of Stone-Wales (SW) defects, i.e., longitudinal and circumferential SW defects, on carbon nanotubes (CNTs) result in two different electronic structures. Based on density functional theory we have shown that the longitudinal SW defects do not open a band gap near the Fermi energy, while a relatively small band gap emerges in tubes with circumferential defects. We argue that the band gap opening in presence of circumferential SW defects is a consequence of long-range symmetry breaking which can spread all the way along the tube. Specifically, the distribution of contracted and stretched bond lengths due to the presence of defects, and hopping energies for low-energy electrons, i.e. the $2p_z$ electrons, show two different patterns for the two types of defects. Interplay between the geometric features and the electronic properties of the tubes have been also studied for different defect concentrations. Considering $\pi$-orbital charge density, it has also been shown that the deviations of bond lengths from their relaxed length result in different ÒdopingÓ for two defect orientations around the defects -Ð electron-rich for circumferential defect, and hole-rich for longitudinal one. We have also shown that in the tubes having both types of defects, circumferential defects would dominate and impose their electronic properties.  
\end{abstract}

\submitto{JPCM}

\maketitle

\section{Introduction}

The unique mechanical, thermodynamic and electronic properties of carbon nanotubes (CNTs), since their discovery \cite{i1}, have nominated them as a building block advanced technologies in future \cite{i2}. As a conductor, a single-wall CNT, depending on its chiral vector, could be metallic or semiconducting \cite{i3}. An ideal metallic CNT can be categorized among the good conductors with the conductance equal to $4e^2/h$. However, like any other quantum conductors, the presence of any kind of disorder significantly changes the conductance of metallic CNTs. The effects of various types of disorder and impurities on the mechanical, thermal and electronic properties of CNTs have been widely investigated both theoretically and experimentally. Among others, the study on the effects caused by the presence of Stone-Wales (SW) defects \cite{i4}, a common structural defect on the surface of CNTs, has been particularly of interest to condensed matter physicists \cite{i5}. A SW defect, which is made by a 90 degree rotation of a single carbon-carbon bond, usually emerges during the growth process of CNTs and other curved graphitic structures, like graphene \cite{i6}. This kind of defect can be introduced to the system, e.g., by single electron irradiation \cite{i7}, or by exerting stress and tension to CNTs \cite{i8}. It is verified that the SW defects are chemically very active \cite{i9} and their presence abruptly changes the mechanical and electronic properties of CNTs \cite{i10}. For example, it has been predicted that the presence of SW defects on an ideal metallic CNT can make it semiconducting. Moreover, it has been shown that the defective semiconducting samples with specific defect concentration can show metallic properties as well \cite{i11}.

Here, we present the results of our first-principle calculations on armchair CNTs containing various SW defect concentrations and two distinct orientations, namely longitudinal (type I), and circumferential (type II). All other orientations will be symmetrically the same as these two. Although some researchers have already addressed the issue of, not yet fully investigated, different orientations of SW defects on infinitely long semiconducting zigzag \cite{i12}, and finite armchair CNTs \cite{i13}, and their induced effects, but, the possible relation between the structural and the electronic properties of CNTs with SW defects is a less studied subject. Our study focuses on different crucial effects brought about by the two different orientations of a SW defect on the geometric and electronic properties of armchair CNTs.

This paper is organized as follows: In Sec. II, we outline the computational method that we have used. Results are presented in Sec. III and the related discussions follow afterwards. Finally, we summarize our results and present our concluding remarks in Sec. IV.

\section{Method of Calculation}

We have used the Kohn-Sham density functional theory (KS-DFT) \cite{m1} to find the relaxed atomic coordinates, and to study the effect of geometrical properties on the electronic structure of armchair carbon nanotubes having SW defects. We have benefited from finite-range pseudo-atomic orbitals implemented in the SIESTA density functional code \cite{m2}, which allowed our computations to be efficiently performed. We have used a Troullier-Martins type pseudopotential \cite{m2-2} for carbon with a double-$\zeta$, singly polarized orbital for the basis set, along with a $1\times1\times18$ Monkhorst-Pack grid \cite{m3} in the reciprocal lattice for both conformal relaxation and electronic structure calculations. We have also employed periodic boundary conditions along the tube axis. 

It is well known that the local density approximation (LDA) for the exchange-correlation energy functional, $E_{XC}$, works well for nearly homogeneous electronic systems, whereas, the generalized gradient approximation (GGA) for $E_{XC}$, in principle, should work better than LDA for systems with considerable electronic density variations. Therefore, for the purpose of studying the possible charge inhomogeneity on the surface of carbon nanotubes in the form of electron- and hole-rich regions, the use of GGA seems to be a more appropriate choice. We have also used Perdew-Burke-Ernzerhof parameterization, which usually gives good results for carbon-based structures in comparison with experimental data \cite{m4}. A variable-cell relaxation has been performed on both perfect and defected CNTs until the maximum force acting on the atoms and the maximum stress tensor element fell below 0.01 eV/$\AA$ and 0.25 GPa, respectively. The target pressure has been set to 0.0 GPa.

In order to calculate the local charge density (CD), we considered the projected density of states, defined as
\begin{equation}
g^{(i,m)}(\epsilon)=\sum_{n} {\delta(\epsilon-\epsilon_n)}
\left|\left<\phi_m^i | \Psi_n\right>\right|^2,
\end{equation}
where $\Psi_n$'s are the Kohn-Sham orbitals, and $\phi_m^i$ is the $m^{th}$ orbital of the $i^{th}$ atom. Therefore, the contribution of the $i^{th}$ atom to the total CD can be written as
\begin{equation}
\rho^{i}=\sum_m{\int_{-\infty }^{E_F } g^{(i,m)}(\epsilon)
f(\epsilon)d\epsilon},
\end{equation}
where $f(\epsilon)$ is the Fermi-Dirac distribution function.\\

\section{Results and Discussion}

Here, we present the results for the effect of two orientations of SW defects, mostly on a (5,5) CNT, which is a metallic one in its perfect form. \\

Figure 1 shows the relaxed configuration of a (5,5) CNT with two types of defects Ð- (a) shows the longitudinal (type I), and (b) shows the circumferential (type II) defect. There are 180 atoms in this sample. 16 atoms with 19 bonds connecting them form a SW defect. The bond lengths in type I defect in (a) are symmetric with respect to the plane passing through the central bond of the SW defect and the tube axis. They are also symmetric with respect to the perpendicular plane passing through the mid-point of the central bond. The bond lengths are also symmetric with respect to the mid-point of the central bond. Type II defect in (b) only shows the point symmetry. Moreover, the bond lengths are relatively longer in the case of type II defect. The obtained bond lengths are in agreement with previous first-principles calculations on such defective systems \cite{i13}\footnote{The naming of longitudinal and circumferential defects in Ref. \cite{i13} is different from this work.}. This difference in the relaxed configuration of the two types of defects signals the possibility of different electronic structures induced by them. 

\begin{figure}
\centerline{
\includegraphics[height=7.5cm]{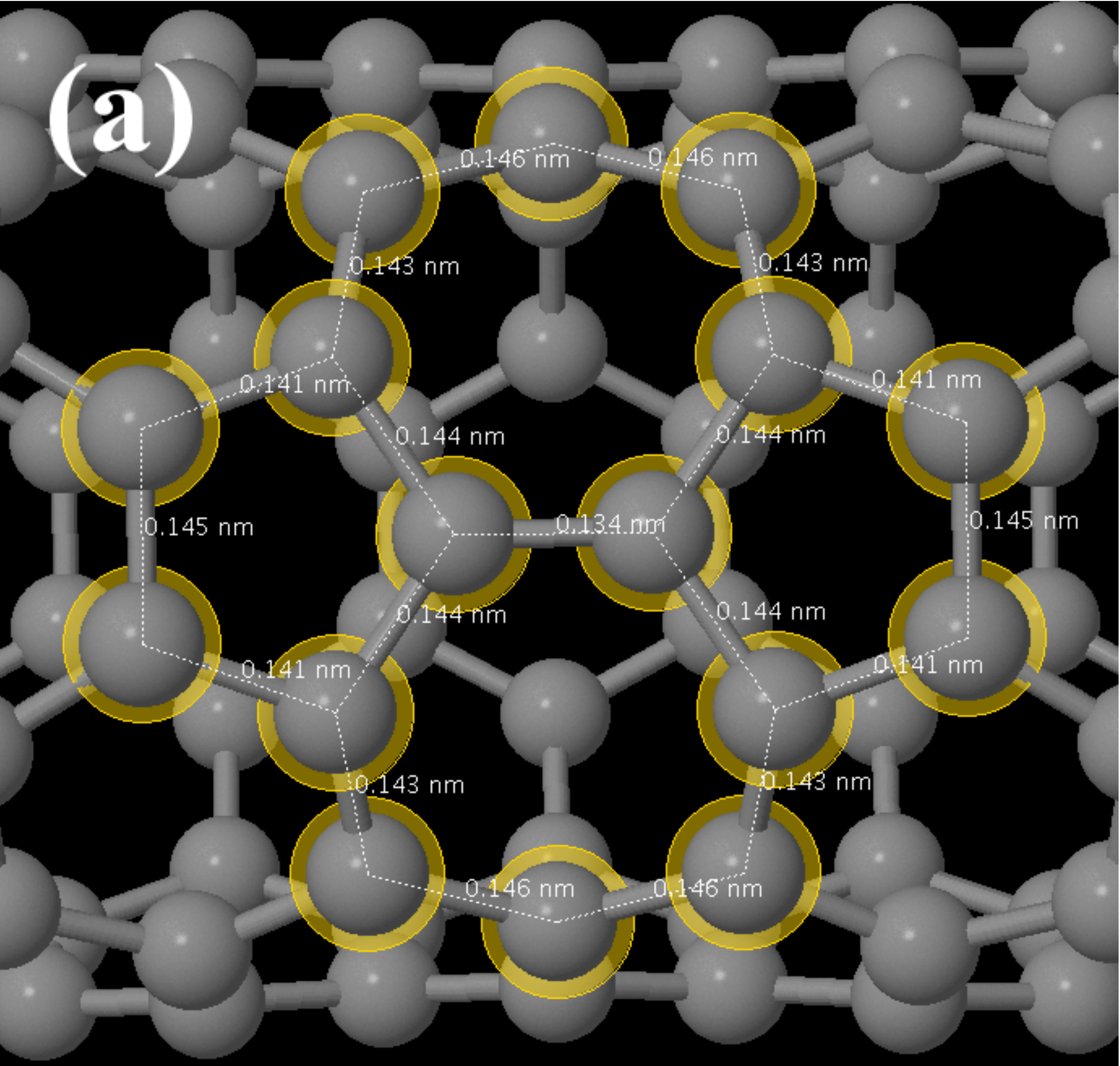} \hspace{0.05cm} 
\includegraphics[height=7.5cm]{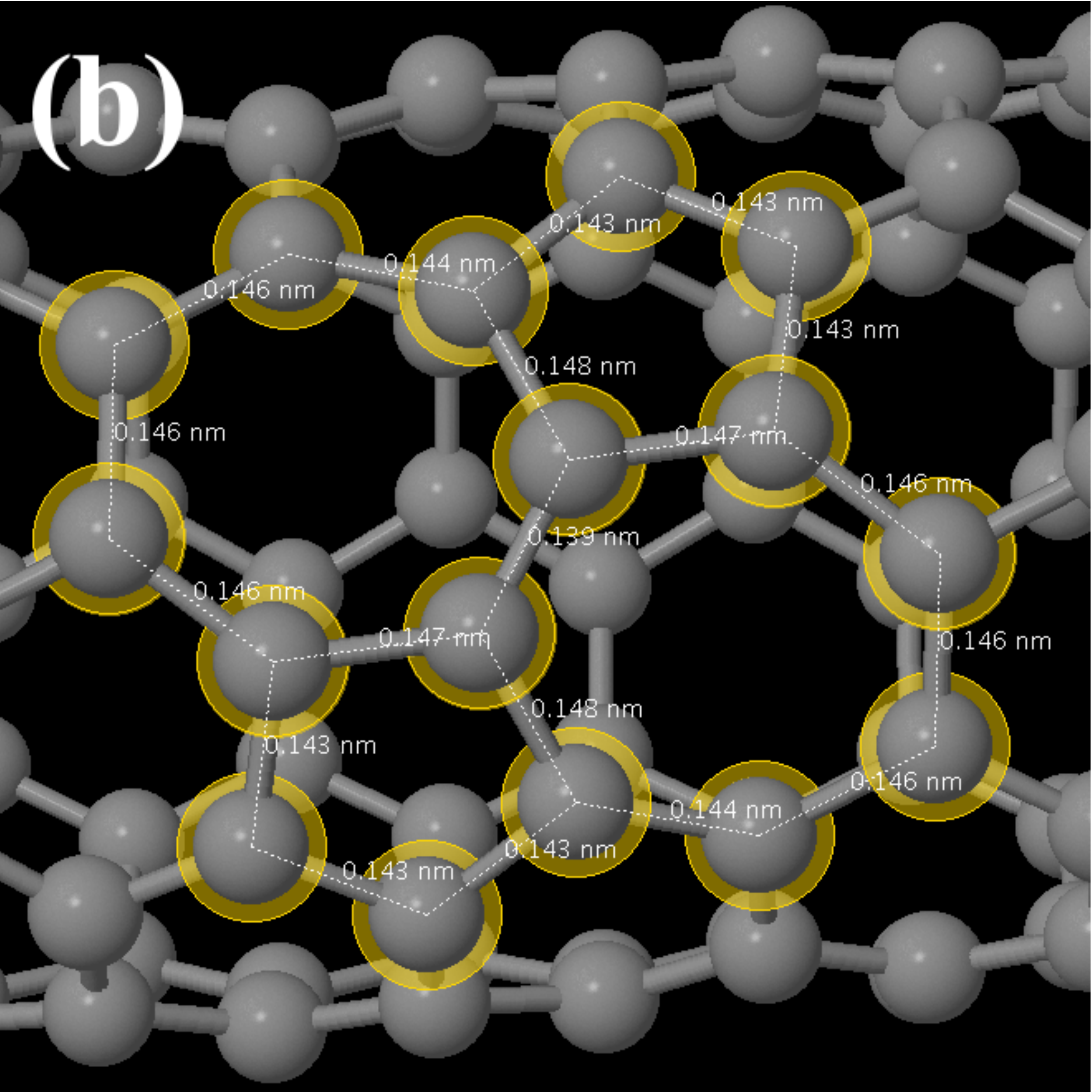} 
}
\caption{Relaxed configuration of two types of SW defects on a (5,5) CNT. The 19 bond lengths of the SW defect are also shown in the figure. In longitudinal defect (type I), the bond length distribution is symmetric with respect to a plane passing through the central bond in the SW defect and the tube axis. The bond lengths are also symmetric with respect to the plane perpendicular to the tube axis passing through the mid-point of the central bond. Moreover, point symmetry is also seen with respect to the mid-point of the central bond in the case of type I defect. In circumferential defect (type II), only the point symmetry is seen.} 
\end{figure}\label{fig1}

Figure 2 shows the electronic density of states (DOS) for two (5,5) CNTs with one SW defect per 180 carbon atoms. Here, we consider one SW defect in the unit cell, so the defect concentration is defined as $1/N$, $N$ being the number of carbon atoms in the unit cell. As such, the defect concentration in the samples shown in figure 2 is 0.55$\%$. Panel (a) shows the DOS for the CNT with type I defect, and (b) shows the same quantity for a CNT with type II. The dashed line in both panels is the DOS for a perfect CNT consisting of the same number of atoms. At Fermi energy, no gap is observed in (a), whereas in (b) an approximately 60 meV gap emerges. The observed difference is quite remarkable because it shows that the same type of topological defect can cause a (semi)metal-semiconductor transition in armchair tubes depending only on the orientation of the defect. In spite of the fact that the conventional density functional theory usually underestimates the semiconducting band gap due to the problems encountered in describing the excited states of the system \cite{m5}, the observed band gap has a considerable value in this specific defect concentration. 

\begin{figure}
\centerline{
\includegraphics[width=9.0cm]{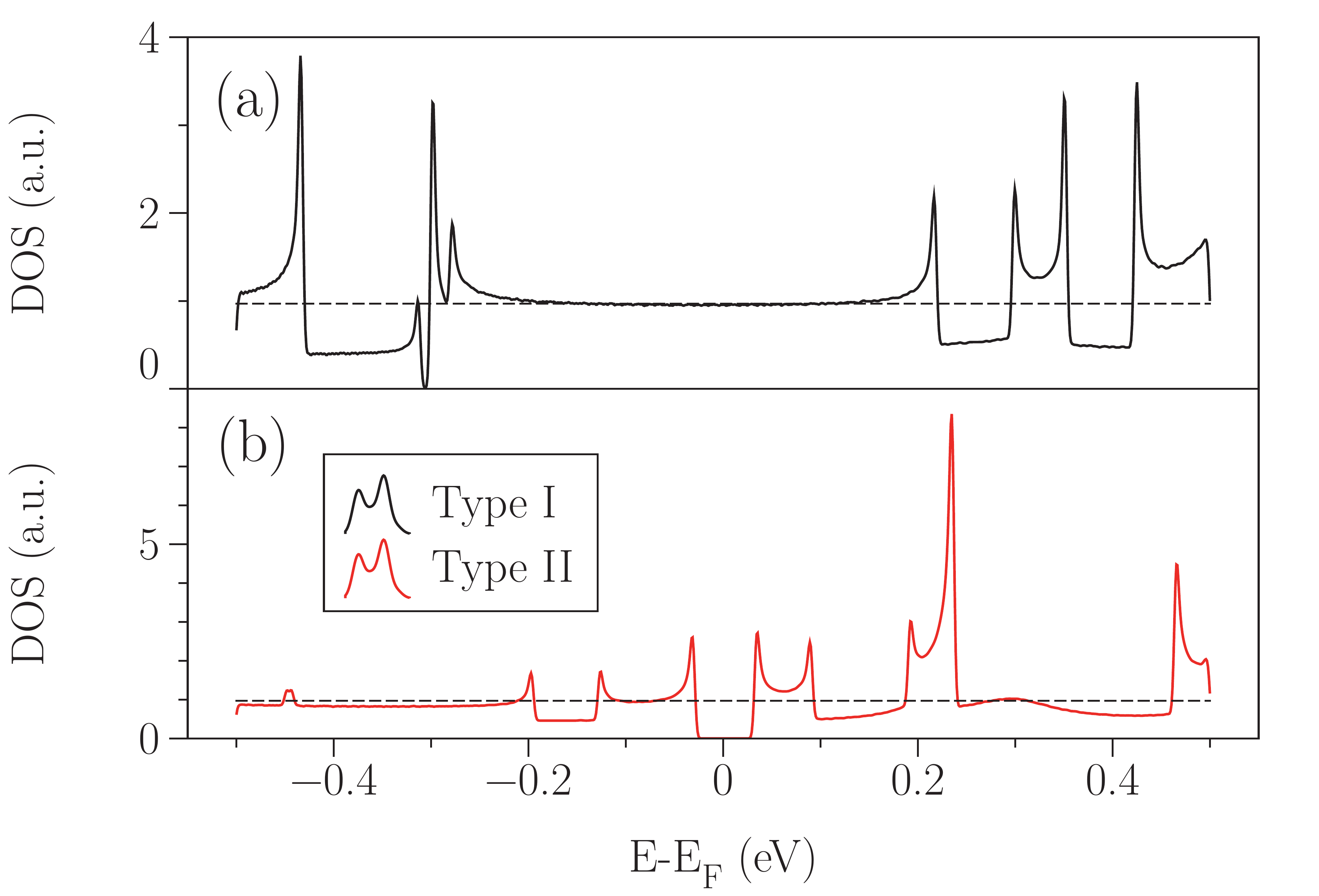}
}
\caption{Electronic density of states near the Fermi energy for a sample consisting of 180 carbon atoms with a type I SW defect (a), and with a type II SW defect (b). The dashed line shows the DOS for a perfect CNT with 180 carbon atoms. In the case of type I defect in (a), no band gap is observed, whereas for the case of type II defect in (b), a gap of $\sim$60 meV can be seen.} 
\end{figure}\label{fig2}

It is worth mentioning here that the above results might seem to be in contradiction to the previously reported results on the energy difference between the HOMO and LUMO states of the tubes with two types of SW defects in Ref. \cite{i13}. However, the systems under consideration in Ref. \cite{i13} are far from periodic, as the authors have considered clusters of defective tubes. This is the reason for the large energy difference, $\sim$5.5 eV, between HOMO and LUMO states in the Ref. \cite{i13}, irrespective of the type of the SW defect. In periodic boundary conditions, which is used in this work, the different induced electronic effects of the two types of defects become clearer. However, the obtained results for the \emph{local} properties of SW defects, e.g., the bond lengths, are in good agreement with each other in both works. Furthermore, it has been shown that the SW defects lying along the tube axis (here, named as type I), have more tendency to narrow down the gap in semiconducting zigzag tubes, compared to those SW defects which stay tilted with respect to the tube axis (here, named as type II) \cite{i12}. Nevertheless, in terms of formation energy, it has been shown that the type II SW defects on armchair tubes act, more or less, the same as the SW defects lying along the axis in zigzag tubes \cite{m5-2}.

Different behaviours are also reflected in the $\pi$-orbital charge density (CD) maps. The corresponding CD distribution on the samples is shown for the two types of defects on $YZ$ and $XZ$ planes in figures 3(a) and 3(b), respectively. $Y$ and $X$ axis, respectively in figures 3(a) and 3(b), are along the diameter of the tubes. In order to find the electron- and hole-rich regions, we subtracted their mean value in each case from the corresponding CD distributions. The black (green) curve in figure 3(a) (figure 3(b)) indicates the contour of zero CD. Figure 3(a) shows the hole-rich region around type I defect, while the region around type II, in figure 3(b), is electron-rich. Also, both electron- and hole-rich regions spread along the tube axis, which is represented as $z$-axis. The same qualitative results were observed also for different defect concentrations.

\begin{figure}
\centerline{
\includegraphics[width=9.0cm]{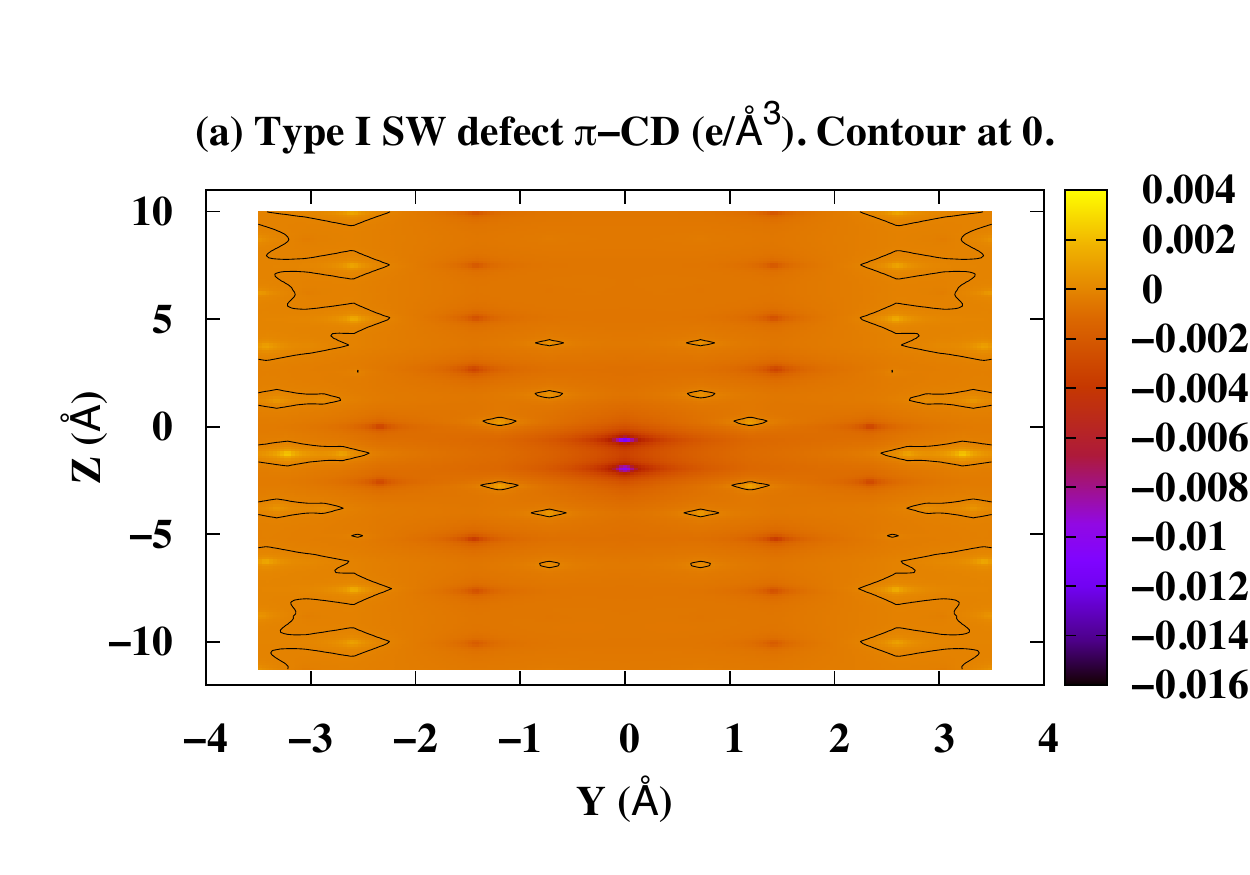}} 
\centerline{
\includegraphics[width=9.0cm]{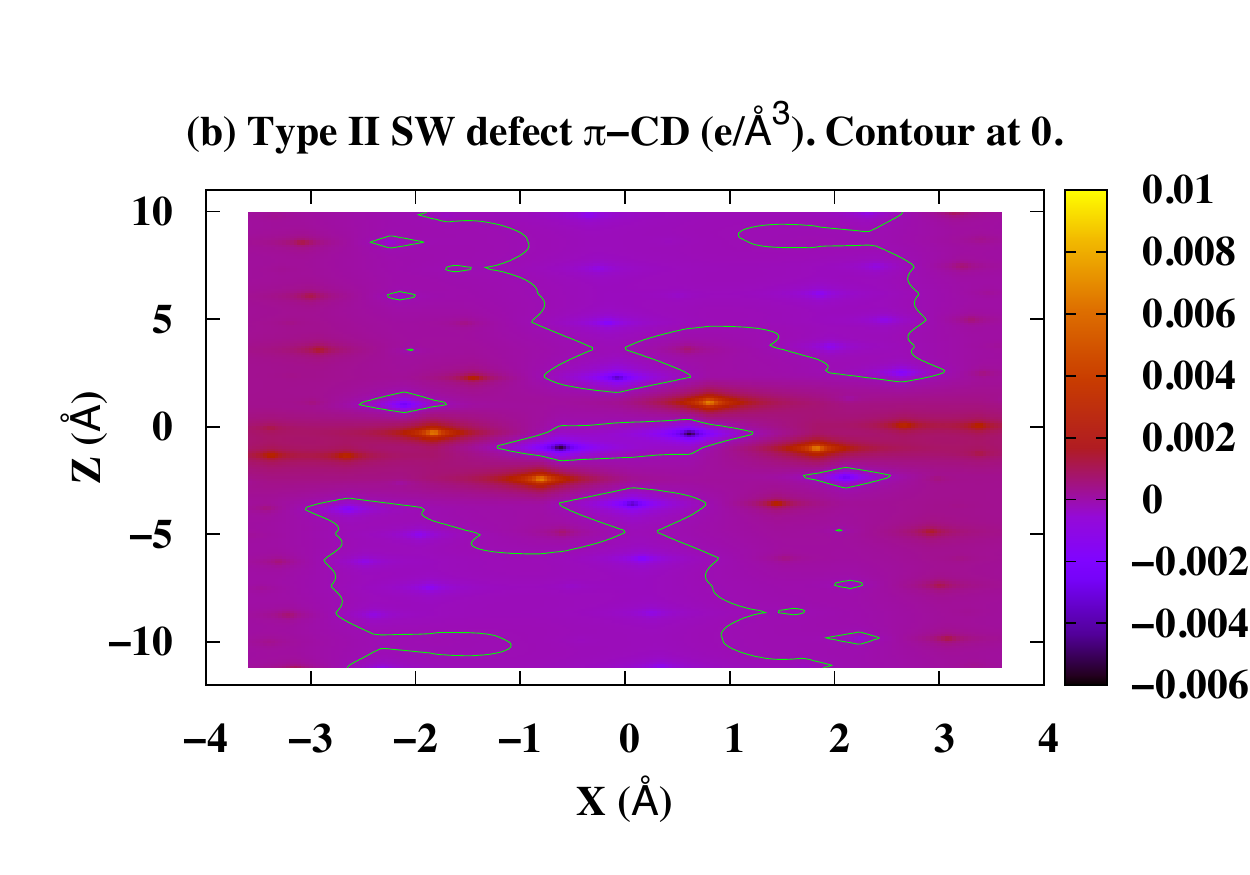} 
}
\caption{Electron- and hole-rich regions for $\pi$-orbitals around two types of SW defect in the sample considered in figure 2. In order to demonstrate electron-rich and electron-deficient regions, the charge density is subtracted from its mean value. The black and green curves, respectively in (a) and (b), indicate the zero charge density. The tubes are placed along $Z$. The CD is plotted on $YZ$ plane in (a) and $XZ$ plane in (b).} 
\end{figure}\label{fig3}

Considering the results in figure 1 and figure 3, the electron-rich (hole-rich) regions have occurred in the regions with relatively longer (shorter) bond lengths. The $\pi$-orbital CD has been extensively studied on corrugated graphene samples \cite{m6}. It has been recently shown that the bond length distribution mainly anti-correlates with $\sigma$-orbital CD, and the $\pi$-orbital CD mainly correlates with the curvature of the surface. High curvature regions are generally regions with longer bond lengths. So, the $\pi$-electron-rich regions on graphene occur on the regions with longer average bond length \cite{m7}. Our findings are consonant with the reported results on graphene. The observed CD inhomogeneity brought about by the SW defects might be a reason for the high reactivity of SW defects compared to the other regions on CNTs \cite{i9}.\\

In order to describe these two different behaviours, we investigated the extent to which the tubes were deformed from their perfect structure in each defect concentration.
\begin{table}
\caption{\label{tab1} Change in the unit cell size along the tube axis. C stands for the SW defect concentration.}
\footnotesize\rm
\begin{tabular*}{\textwidth}{@{}l*{15}{@{\extracolsep{0pt plus12pt}}l}}
\br
SW type&C=0.71\%&C=0.55\%&C=0.50\%&C=0.35\%\\
\mr
Type I&0.8\%&0.7\%&0.6\%&0.4\%\\
Type II&-0.3\%&-0.3\%&-0.2\%&-0.2\%\\
\br
\end{tabular*}
\end{table}
Table 1 shows the change in the unit cell size along the tube axis with respect to the perfect sample in each concentration. Although the change is small in both types, but different defect types behave differently. The unit cell along the tube axis, compared to that in the perfect sample, becomes a bit longer in the case of samples with type I defects, while it becomes a bit shorter in the case of type II defects. The change becomes smaller, in both cases, as the SW defect concentration is lowered.

Moreover, the deformation was studied in terms of average displacements of the carbon atoms in the defected systems from their relaxed positions in the perfect sample, as a function of defect concentration (see figure 4(a)). In this calculation, the displacement of the two atoms corresponding to the central bond of the SW defect is neglected, because as a starting configuration of a SW defect in the relaxation process, that bond was manually rotated to form two heptagons and two pentagons. In figure 4(a) type II defects show an overall larger average displacement compared to type I. 
Figure 4(b), shows the total energy difference between the samples with two types of defects, as a function of SW defect concentration. The formation of type II defect seems to be more favorable in the range of considered concentrations. This is in agreement with previous tight-binding and first-principles calculations on armchair CNTs with two different SW defect orientations \cite{i13,m5-2}. The differences between type I and type II defects, shown in figure 4, are due to the fact that there is a remarkable difference between elastic moduli of CNTs in longitudinal and circumferential directions \cite{r1}. Descending behaviour, on the other hand, suggests that for very low concentrations the effect of two defects on the geometrical and the electronic properties of the tubes should be local and, more or less, the same.
\begin{figure}
\centerline{
\includegraphics[width=9.0cm]{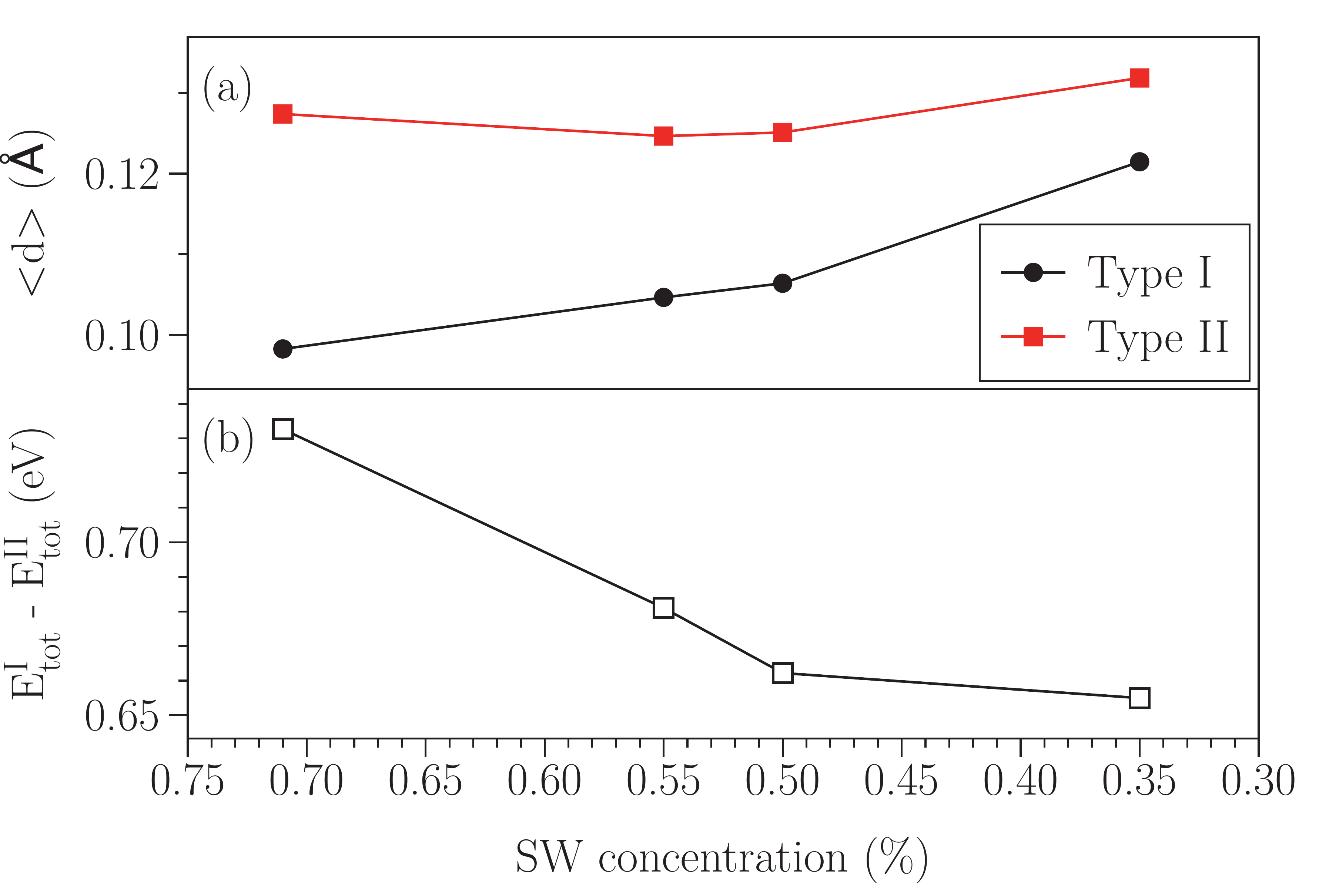}
}
\caption{(a) The average atomic displacements, from the perfect structure, due to the presence of the two types of SW defects on a (5,5) tube as a function of defect concentration (the horizontal axis is reversely plotted; lower concentrations correspond to longer samples). The concentration is defined as the number of SW defects divided by the number of carbon atoms in the unit cell. (b) The total energy difference in the tubes with two types of defects, as a function of defect concentration.} 
\end{figure}\label{fig4}
In the electronic structure calculations on the samples represented in figure 4, no gap was observed in CNTs with type I defects, irrespective of the defect concentration. However, the magnitude of the band gap  in type II samples varies as a function of defect concentration.

Figure 5 shows (a) the area per SW defect, (b) the Fermi energy, and (c) the band gap, as a function of defect concentration in CNTs with type II defects. The overall descending behaviour of the band gap, as well as that of the Fermi energy, can be seen in figure 5(b) and figure 5(c). The variations of the band gap and the Fermi energy qualitatively appear to agree with the variations of the area per SW defect which can be considered as yet another factor by which the deformation of the tube can be characterized.
\begin{figure}
\centerline{
\includegraphics[width=9.0cm]{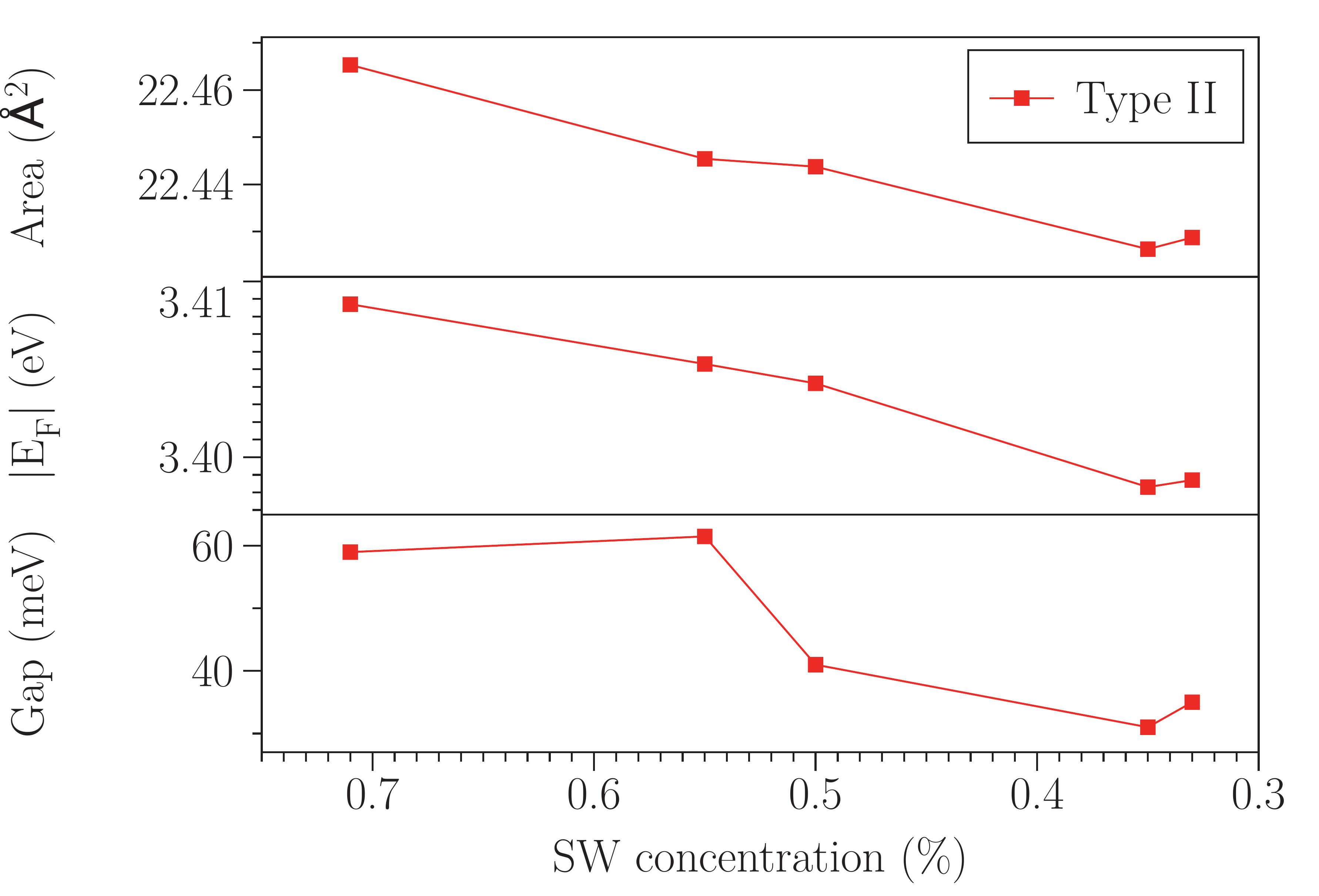}
}
\caption{(a) Area per SW defect, (b) absolute value of the Fermi energy, and (c) band gap of (5,5) CNTs with type II SW defects, as a function of defect concentration.} 
\end{figure}\label{fig5}
Apart from the systematic error that usually happens in the calculation of the band gap within the framework of the conventional density functional theory, the interplay between the geometrical features and the electronic properties of the defected CNTs is clear. The presence of a SW defect also makes the tube radius have a range of variation along the tube, instead of having one value throughout. However, no explicit correlation between the electronic properties, i.e. the Fermi energy as well as the band gap of the tubes and their radius range in different defect concentrations was observed.

The gap opening can be attributed to the fact that SW defects can form a long-range deformation, which spreads from one side of the tube to the other. This fact is demonstrated in figure 6, which shows the average bond length map on (5,5) tubes with two type of SW defects and different concentrations. The bond lengths shorter (longer) than the calculated value for the perfect tube, 1.436 $\AA$ which is in agreement with other calculations based on DFT \cite{r2}, are shown in dark (light) spheres. In the case of type I defects, figure 6(a), the dark spheres form a path from one side of the sample to the other in all defect concentrations. In the case of type II defects, figure 6(b), it is the stretched bonds that form the path. Contracted bonds on the tubes with type I defects, lead to relatively larger hopping energies (and overlaps) for $\pi$-orbitals, and in turn result in low-energy conducting channels spreading along the tube. These electrons, with even higher hopping energies than the conducting perfect sample, can still easily pass through the tube when an external field is applied. On the other hand, in type II case, regions with shorter bonds (higher hopping energies and overlaps) form isolated Òislands,Ó which are separated from each other, and therefore, no conducting channels will emerge. In regions with longer bonds, which spread along the tube in the case of type II defects, the hopping energies become small, and therefore, hopping would be an unlikely process. Consequently, in agreement with the calculated DOS shown in figure 2, a relatively small band gap in CNTs with type II defects would emerge.

\begin{figure*}
\centerline{
\includegraphics[width=6.0cm]{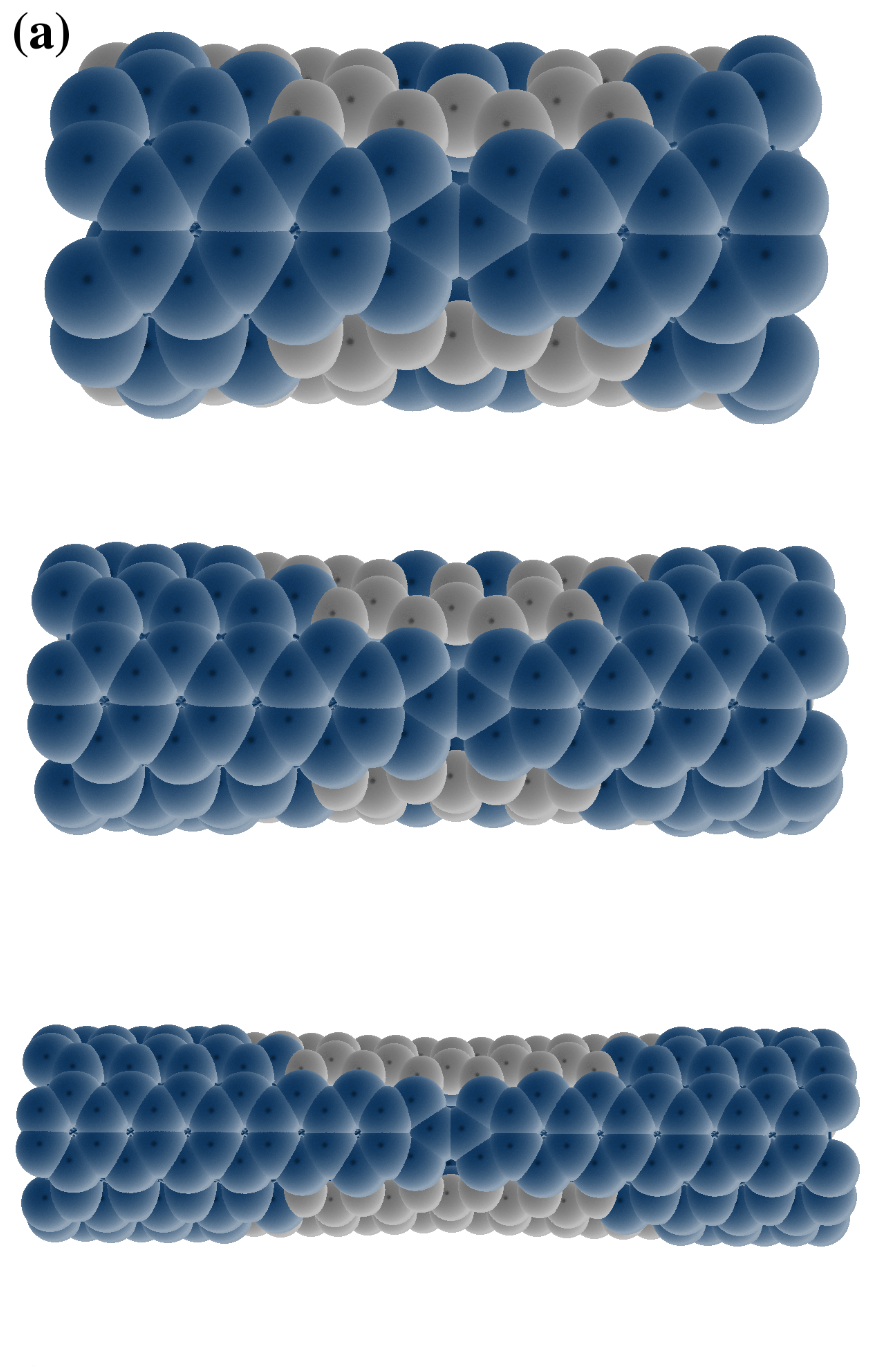} 
\includegraphics[width=6.0cm]{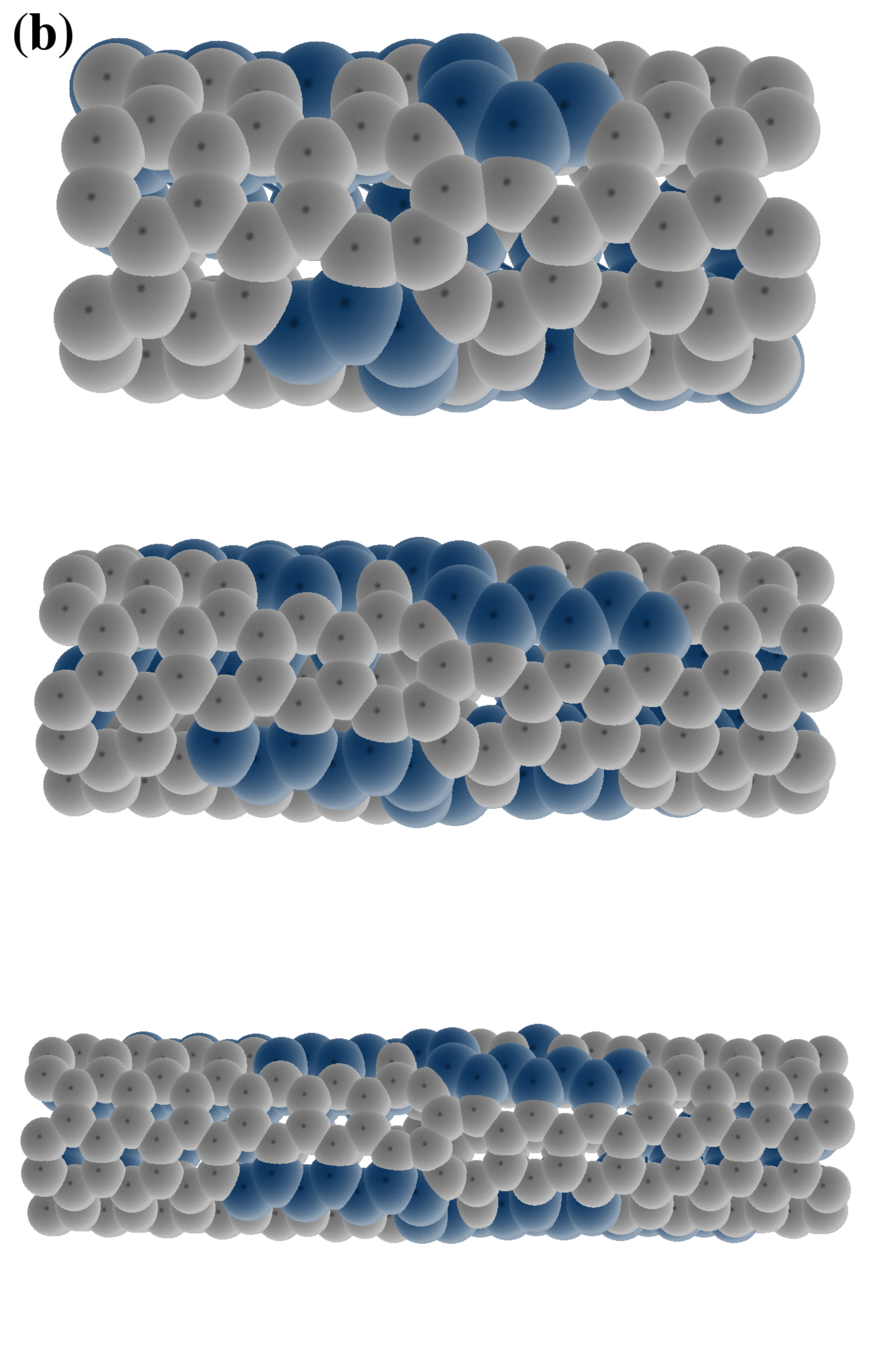} 
}
\caption{Contracted (dark spheres) and stretched (light spheres) average bond lengths, with respect to the calculated value for a perfect (5,5) tube having bond length of 1.436 $\AA$, for two types of SW defects -- type I is shown in panel (a), and type II is shown in panel (b). From top to bottom, SW concentration varies from 0.71$\%$ (upper panel) to 0.50$\%$ (middle panel) to 0.36$\%$ (lower panel). Both contracted and stretched bonds form a path in all concentrations, which spreads from one side of the tube to the other. Contracted (stretched) bonds correspond to higher (lower) hopping energies between neighboring atoms.} 
\end{figure*}\label{fig6}

This fact is shown in figure 7 where the hopping energies between $\pi$-orbitals of the nearest neighbors on the paths shown in figure 6 are plotted as a function of position along the tube axis. The calculation is done for the sample with 180 carbon atoms and the values are subtracted from the same hopping energies in the perfect sample.
\begin{figure}
\centerline{
\includegraphics[width=9.0cm]{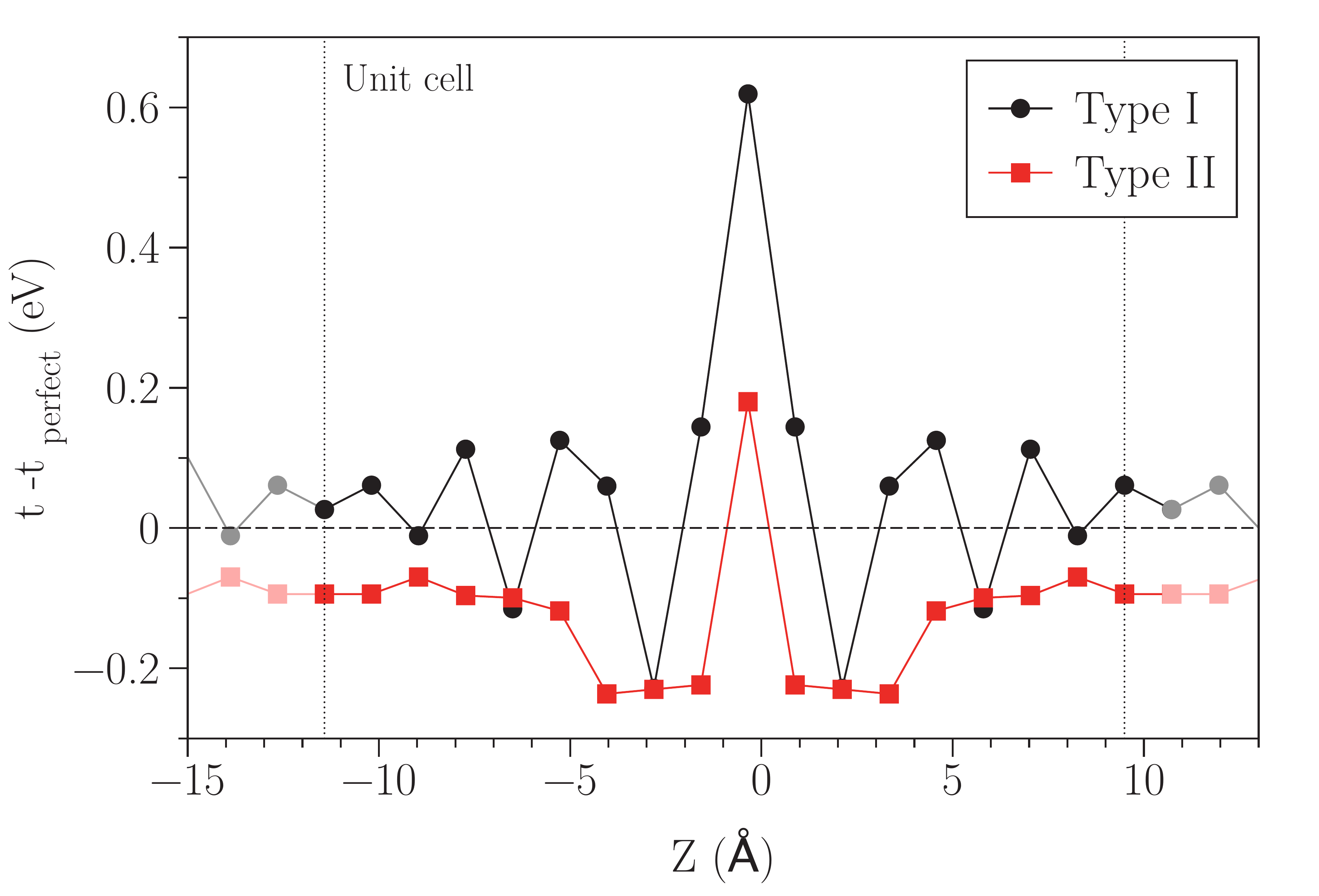}
}
\caption{The hopping energy between the $\pi$-orbitals of the nearest neighbors in a (5,5) tube consisting of 180 atoms with two different defect types, along the paths shown in figure 6. The hopping energies for CNT with type I (type II) defect are shown in black circles (red squares). Moreover, the hopping energies are subtracted from the hopping energy calculated for the perfect sample.} 
\end{figure}\label{fig7}
As is clear, the hopping energies in the case of type II defects are considerably smaller than the  calculated value for the perfect sample, whereas, the hopping energies for the case of type I defects are greater than those in the perfect sample. 

Furthermore, the change in the hopping energies along the tube circumference, from its value in perfect sample, should be lower in tubes with longer diameter. In other words, the amount of scattering of the electronic wave functions from the defect, on average, is lower for the tubes with longer diameters.
\begin{figure}
\centerline{
\includegraphics[width=9.0cm]{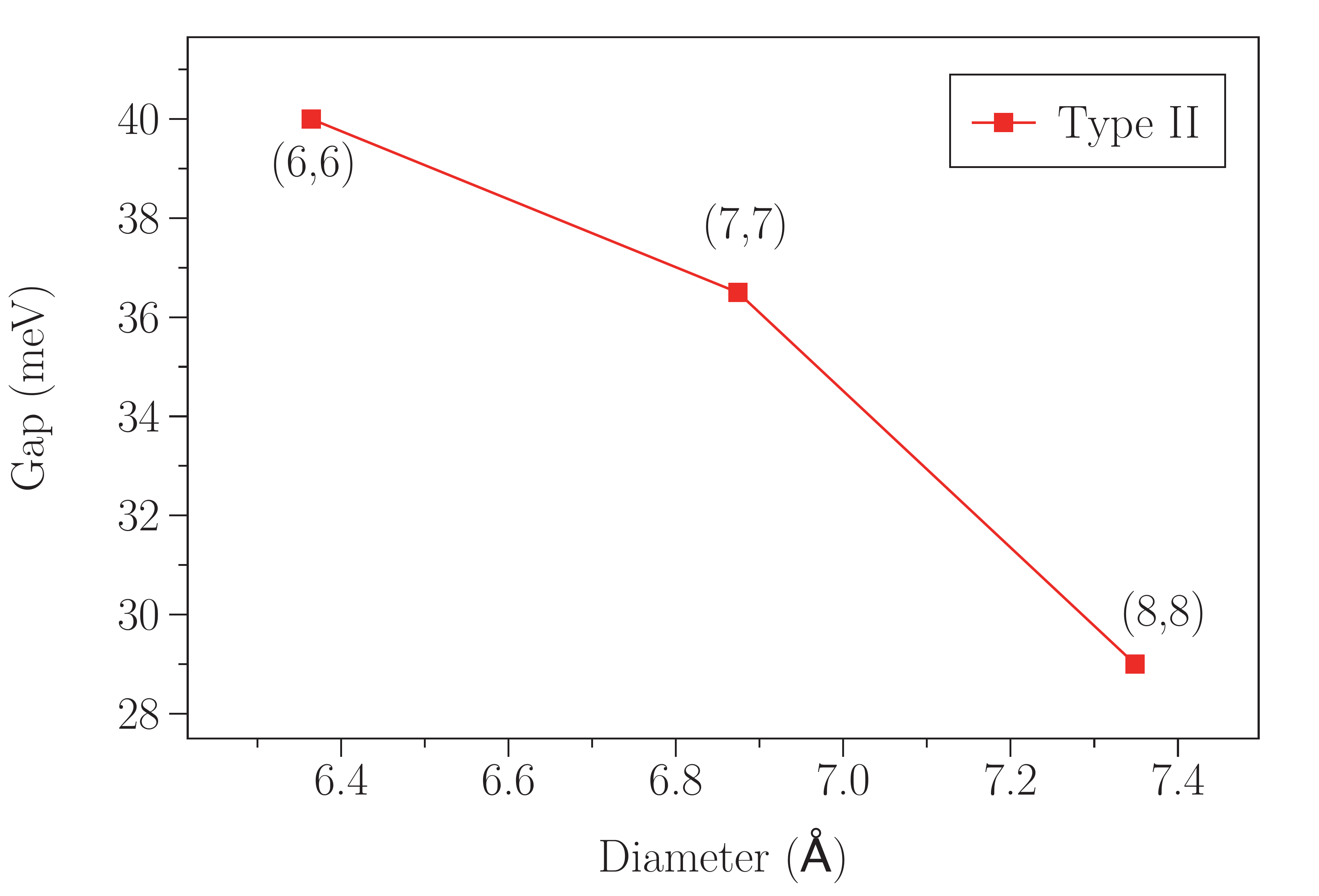}
}
\caption{Band gap of the tubes with type II defects as a function of their diameter. For all of the samples, one type II SW defect in five translational period along the tube axis is considered, giving rise to one defect per 120, 140 and 160 carbon atoms for (6,6), (7,7) and (8,8) CNTs, respectively.} 
\end{figure}\label{fig8}
Figure 8 shows the band gap as a function of the tube diameter. Descending behaviour is presumably due to the fact that the degree of deformation along the tube circumference is lower in tubes with longer diameters. In tubes with type I defects, with the same diameter and defect concentrations as in figure 8, no gap was observed, as expected.

Finally, we studied a (5,5) tube consisting of 300 atoms in the unit cell and having both types of SW defects. The defects are relatively along a line, as shown in figure 9(a). In figure 9(b), we show the DOS of the sample appearing in figure 9(a). Also shown in figure 9(b), is the DOS of a CNT with 300 atoms having only type II defects.
\begin{figure}
\centerline{
\includegraphics[width=10.0cm]{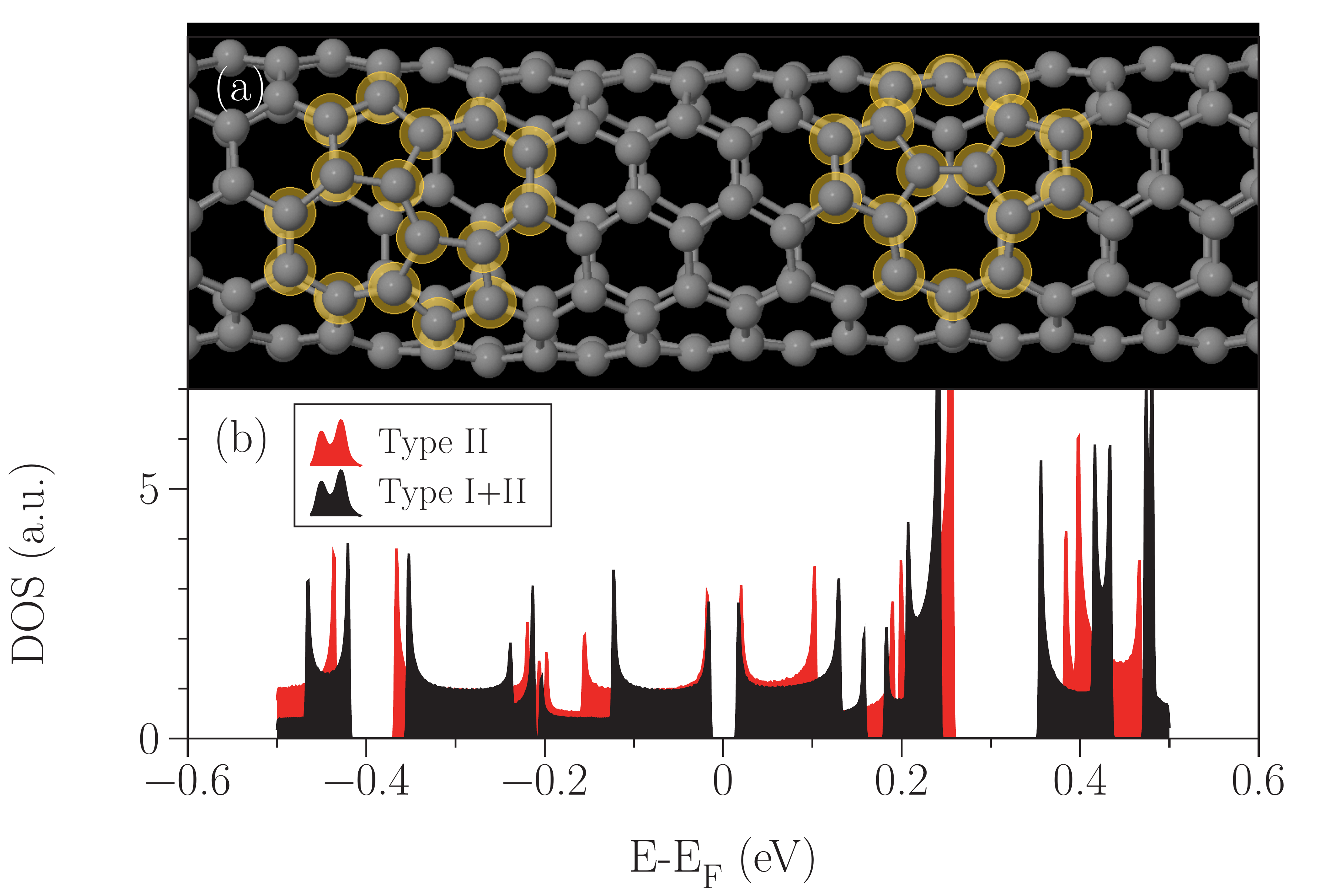}
}
\caption{Relaxed coordinates (a) and the band structure (b) of a (5,5) tube having both types of SW defects (black), and only type II defects (red).} 
\end{figure}\label{fig9}
The present configuration for the defects was chosen to find out which defect type dominates in imposing its effect in terms of band gap opening. The paths with contracted and stretched bond lengths, which were shown to form individually on the CNTs with only one type of defects, intersect with each other in the present configuration. As is clear in figure 9(b), the type II defects have stronger effects which results the gap opening. The gap around Fermi energy is almost the same in samples with only type II defects and the systems having both types of defects. There are also two relatively bigger gaps around -0.4 eV and 0.3 eV. As is shown in red colours, these gaps are also present in the DOS of the CNT with only type II defects. These results also suggest that in experimental studies on SW defected CNTs, it is likely to observe a band gap near the Fermi energy, since formation of a type II SW defect, which is energetically favorable compared to a type I defect, will introduce a band gap, irrespective of the concentration of type I defects on the tube.\\

\section{Conclusions}

We have studied the effect of two different orientations of the Stone-Wales defect and the induced geometrical deformations on the electronic properties of armchair carbon nanotubes, using Kohn-Sham density functional theory. It has been shown that in the case of circumferential defects, a small band gap would emerge due to the elongated bond lengths, and lower hopping energies between neighboring atoms. However, in the case of longitudinal defects, no band gap was observed, no matter what the defect concentration was. Moreover, it has been shown that in this case, nearest neighbor hopping energies, along a path from one side of the sample to the other covering the defect itself, are relatively larger than those in the perfect sample. Also, it has been shown that two types of defects result in two different electronic structure, with either electron-rich or hole-rich regions around them. Furthermore, in the tube with both types of SW defects, the circumferential defects dominate in imposing their effect in terms of band gap opening. This is due to the fact that the degree of geometrical deformation along the circumference is higher than that in the longitudinal direction. The most important point presented here is that a single topological defect, which is imagined as a rotation of a carbon-carbon bond by 90 degrees, can cause a (semi)metal-semiconductor transition in metallic carbon nanotubes depending on its orientation with respect to the tube axis. \\

\ack{
P.P.-A would like to thank S. Panahian, for very helpful remarks on the paper. P.P.-A also wants to acknowledge N. Abedpour and T. Jadidi for fruitful discussions, N. Nafari for critically reviewing the manuscript, and M. R. Rahimi Tabar and P. Maass for providing high-performance computing facilities.
}

\end{document}